\documentclass[12pt,showpacs,preprintnumbers,floatfix]{article}
\usepackage{graphicx}
\usepackage{dcolumn}
\usepackage{bm}
\usepackage{amsfonts}
\usepackage{amsthm}
\usepackage{amsmath}
\usepackage{amssymb}
\usepackage{epsfig}
\usepackage{cite}

\def\be{\begin{equation}}
\def\ee{\end{equation}}
\def\ba{\begin{eqnarray}}
\def\ea{\end{eqnarray}}

\def\bdm{\begin{displaymath}}
\def\edm{\end{displaymath}}

\def\ga{~\mbox{\raisebox{-.6ex}{$\stackrel{>}{\sim}$}}~}
\def\bq{\begin{quote}}
\def\eq{\end{quote}}

 at 10truept

\newcommand{\bea}{\begin{eqnarray}}
\newcommand{\eea}{\end{eqnarray}}

\newcommand{\half}{\frac{1}{2}}
\newcommand{\bi}{\begin{itemize}}
\newcommand{\ei}{\end{itemize}}

\newcommand{\pd}{\dot{\phi}}
\newcommand{\pds}{\dot{\phi}^2}

\newcommand{\vd}{\dot{\varphi}}
\newcommand{\vdd}{\ddot{\varphi}}
\newcommand{\ld}{\dot{\lambda}}
\newcommand{\ldd}{\ddot{\lambda}}
\newcommand{\lds}{\dot{\lambda}^2}

\newcommand{\vp}{{\varphi}^{\prime }}
\newcommand{\vpd}{\dot{\varphi}}
\newcommand{\vps}{{\varphi}^{\prime \; 2}}
\newcommand{\vpp}{{\varphi}^{\prime \prime}}
\newcommand{\lp}{{\lambda}^{\prime}}
\newcommand{\lps}{{\lambda}^{\prime \; 2}}
\newcommand{\lpp}{{\lambda}^{\prime \prime}}

\newcommand{\dlm}{\Delta_\phi {\cal L}_m}

\newcommand{\beq}{\begin{equation}}
\newcommand{\eeq}{\end{equation}}
\newcommand{\beqa}{\begin{eqnarray}}
\newcommand{\eeqa}{\end{eqnarray}}

\def\ga{~\mbox{\raisebox{-.6ex}{$\stackrel{>}{\sim}$}}~}

\def\ltap{\ \raise.3ex\hbox{$<$\kern-.75em\lower1ex\hbox{$\sim$}}\ }
\def\gtap{\ \raise.3ex\hbox{$>$\kern-.75em\lower1ex\hbox{$\sim$}}\ }
\def\gl{\ \raise.5ex\hbox{$>$}\kern-.8em\lower.5ex\hbox{$<$}\ }
\def\roughly#1{\raise.3ex\hbox{$#1$\kern-.75em\lower1ex\hbox{$\sim$}}}

\begin{document}

\thispagestyle{empty}
\begin{flushright}
arXiv:0712.1820 [hep-th]\\
December 2007
\end{flushright}
\vspace*{1.2cm}
\begin{center}
{\Large \bf Geometric Precipices in String Cosmology}\\

\vspace*{1.2cm} {\large Nemanja Kaloper$^{a,}$\footnote{\tt
kaloper@physics.ucdavis.edu}, and Scott Watson$^{b,}$\footnote{\tt
watsongs@umich.edu} }\\
\vspace{.5cm} {\em $^a$Department of Physics, University of
California, Davis, CA 95616}\\
\vspace{.2cm} {\em $^b$Michigan Center for Theoretical Physics,
University of Michigan,
Ann Arbor, MI 48109}\\

\vspace{1cm} ABSTRACT
\end{center}
We consider the effects of 
graviton multiplet fields on transitions between 
string
gas phases. Focusing on the dilaton field, we show that it
may obstruct transitions between different
thermodynamic phases of the string gas, because the sign of its 
dimensionally reduced, $T$-duality invariant, part is conserved when the energy density of
the universe is positive. Thus, many interesting solutions 
for which this sign is positive end up in a future curvature singularity.
Because of this, some of the thermodynamic
phases of the usual gravitating string gases behave like
superselection sectors. For example, a past-regular 
Hagedorn phase and an expanding FRW phase dominated by string
momentum modes cannot be smoothly connected in the 
framework of string cosmology with positive sources. 
The singularity separates them like
a geometric precipice in the moduli space, preventing the dynamics of
the theory from bridging across.
Sources which simultaneously violate the positivity of energy and 
NEC could modify these conclusions. We 
provide a quantitative measure of positivity of energy and NEC violations
that would be necessary for such transitions. These effects
must dominate the universe at the
moment of transition, altering the standard gas pictures. At present, it is not known how to construct 
such sources from first principles in string theory.

\vfill \setcounter{page}{0} \setcounter{footnote}{0}
\newpage

\section{Introduction}

String theory differs dramatically from quantum field theory at
short distances and high energies. At those scales the extended
nature of strings comes into play, and may presumably regulate
ultraviolet divergences inherent in field theory. Yet, since the
string scale is high, $\ell_{s}{}^{-1} \ga {\rm TeV}$ (with the
bound being saturated only in most optimistic cases), it is well
neigh impossible to test string theory in low energy experiments.
On the other hand, successes of modern cosmology have led to the
picture of an expanding universe, which was much denser and hotter in
its early age. At such high energies, full blown string dynamics
may have taken place. Its effects may be important for many
reasons. In the limits of string theory which are under full
computational control, calculability requires the presence of
additional dimensions of space, which somehow must have ended up
hidden from low energy probes in the course of cosmic evolution.
String theory may also have played a role in controlling and
selecting cosmological initial conditions and subsequent evolution
which produced a large and smooth universe, by inflation or
otherwise. In the course of this evolution, some signatures of
string theory may have survived the long era of cosmic dilution,
and they could yield new insights in the microphysics of our
world. Ultimately, one may hope that string cosmology can shed new
light on cosmological singularities, where field theory
formulations of matter and gravity fail completely, having no
direct means to describe the universe at the densest state.

For these reasons, significant efforts have been invested in the
formulation of string cosmology. Interesting models emerged, which
treat string matter in thermal equilibrium \cite{kripe,brava,tseva}, evolving under the
influence of gravity but with specific stringy phenomena accounted
for in the description of the string matter. This approximation
clearly cannot fully reveal the details of the microscopic
dynamics. Yet, one may at least explore the effects of long range
string phenomena on some of the outstanding cosmological problems \cite{pbb,edd,Battefeld:2005av,eva,colawa,egrja,break,Watson:2004aq,Cremonini:2006sx,Greene:2007sa,Battefeld:2004xw,Bassett:2003ck,Borunda:2006fx}.
An intriguing idea was that at very high energies, because strings
are extended objects, the string matter can undergo a Hagedorn
transition, where the energy density saturates at a finite value
controlled by the Hagedorn temperature, taming the violent
divergences in the field theoretic models of matter \cite{tseva}. If this
occurs, then using the network of string dualities one might be
able to describe the universe beyond the string scale by some dual
picture, which admits a different, but controllable perturbative
theory. 

On the other hand, following the growth of the stringy Hubble
volume, one may imagine that as the universe cools the Hagedorn phase should connect
onto the standard expanding cosmology describing the evolution of
dilute strings. The post-Hagedorn stage should initially be
dominated by relativistic strings, behaving as a radiation
dominated FRW universe \cite{tseva}. Similarly, one can also imagine how the
Hagdorn phase could be an end product of a collapsing universe
controlled by string winding modes which are getting dense and
transmuting into the Hagedorn phase. If these transitions can be
realized consistently, one could explore their
implications\footnote{Some novel ideas for generating
scale-invariant perturbations of the geometry were proposed \cite{ali}-\cite{braplus}, based
on the proposed transitions between different stages of the string
gas. However to date they have not succeeded \cite{kklm}. Perhaps it is
interesting to explore them further, at least to verify how robust
the connection between scale-invariant perturbations and inflation
is.} for large universes such as our own. The question is whether
such transitions are possible.

Our aim here is to study in more detail the effects of the
graviton multiplet fields on the transitions between the string
gas phases. We will focus on the dilaton field, and show that in
fact it stands in the way of smoothly connecting different
thermodynamic phases of the string gas. This is because there is a
discrete charge characterizing the dilaton sector: the sign of the
dimensionally reduced effective dilaton, ${\tt sgn}(\dot
\varphi_s)$ is a conserved quantity, when the energy density of
the universe is non-negative \cite{kalmad}. The solutions fall into branches
characterized by this sign, and have a general feature that the
location of the cosmological singularity strictly correlates with
the sign. On the $(+)$ branch, the singularity is in the future,
while on the $(-)$ branch it is in the past. The
singularity may be either a pole super-inflation, or a Big Crunch,
which can be reinterpreted as a pole super-inflation in the
$T$-dual geometry. On the other hand, if
the effective energy density is allowed to dip below zero
temporarily, the transitions may be facilitated \cite{bruve}, and for
conventional sources which satisfy the null energy condition (NEC)
those that can complete are the $(-) \to (+)$ transitions \cite{kalmad}. Such
solutions describe the cosmologies which develop future
singularities induced by the transitions. The transitions in the
opposite direction, from $(+)$ to $(-)$ branch, are impossible
unless NEC is violated \cite{kalmad,brumad}. This means, that the universes which start
evolving towards a future singularity cannot evade this fate 
as long as NEC holds.

This implies that some of the proposed connections between the
Hagedorn phase and the expanding FRW phase dominated by string
momentum mode radiation \cite{tseva} are impossible in the conventional
framework of string cosmology, since those solutions lie on
different branches. Instead, the decaying Hagedorn stage ends up
as a collapsing FRW cosmology, which crunches into the future
spacelike singularity. Similarly, we will see that a collapsing
universe dominated by string winding modes will not condense into
the long-lived Hagedorn stage evolving towards the attractor,
again because its dilaton charge is different. The thermodynamic
phases of the usual gravitating string gases behave like
superselection sectors, characterized by whether they evolve
towards or away from a singular region, which separates them like
a geometric precipice in the moduli space, where the dynamics of
the theory cannot bring them together.

Sources which simultaneously violate the positivity of energy and 
NEC could modify these conclusions \cite{kalmad,brumad}. There
are ways to mimic violations of positivity of energy in string
theory, such as by inducing negative cosmological constant,
allowing for Casimir energy effects or dimensionally reducing on
compact spaces of positive curvature \cite{brumad}. To evaluate the possibilities and
impossibilities of any NEC violations, we retain the perfect fluid
description and provide a quantitative measure of positivity of energy and NEC violations
that could facilitate the transitions between different
thermodynamic phases of string gas. These effects
must be dramatic, as they need to dominate the universe at the
moment of transition, altering the standard gas pictures \cite{bimaz}. Such
dramatic NEC violations are not derived from first principles at
present. Further, generically there
are dangers that covariant theories which violate positivity of energy and 
NEC may give rise to ghosts. Perhaps it is possible to tame the ghost instabilities
if Lorentz symmetry is broken \cite{nima}, and induce cosmological bounces
away from the singularity as in \cite{markus,paolo}. On the other hand, string
theory in its standard formulation is known not to have ghosts
\cite{stelle,zwie,nimauv}. One must properly account for this when attempting to
build cosmological scenarios which strive to employ string matter
phase transitions as a method for generating cosmological
perturbations. Before inventing mechanisms that rely on crossing
between various phases of string cosmology, one must show that the
bridges that need to be crossed aren't too far, but are firmly
grounded in theoretical foundation.

\section{A Review of Gravitating String Ensembles}

Cosmological solutions notoriously involve spacelike
singularities, where the curvature blows up, and the effective
field theory description of geometry breaks down. In string
cosmology, one hopes that the singularities of the effective
description could be ameliorated by string effects. The
complexities of string theory in time dependent backgrounds are a
formidable hurdle to this program, having so far obstructed
singularity resolutions, even casting doubt on their existence. To
gain some insight in what may be going on, one may instead seek to
improve the effective theory description by resorting to some
fundamental features of string theory, such as dualities, which
may even survive supersymmetry breaking. Indeed, in the string gas
approach one should follow the effective geometry to near the
singularity, and then invoke a hypothesis that as the original
geometry begins to explore curvatures beyond the cutoff, one may
replace it by a dual geometry, where the curvatures again turn up
small and a perturbative description is restored \cite{brava,tseva}. This very
appealing idea thus offers a way to address singularity
resolutions entirely within the effective theory language. In this
framework one can ask questions about the consistency of
duality-related solutions with the desired asymptotic limits to
see if they are at least compatible with the standard lore of
string theory.

This is the philosophy which we will pursue in what follows.
Assuming that we begin in some large universe where effective
theory works well, one can coarse-grain the sources of gravity
that inhabit the geometry, using the rules of string
thermodynamics to determine the leading-order equations of state
and gravitational properties of the string ensembles. This yields
the evolutionary laws for homogeneous cosmologies, which we can
follow to near the strong coupling/large curvature regime, where
the string apparent horizon becomes comparable to the string scale
$\ell_{s} \sim M_{s}{}^{-1}$. At this stage, the interstitial
distance between the microscopic strings becomes small and the
coarse-grained description of cosmology must fail. However, if we
follow the dictum of dualities and take it that as distances
become smaller, there exists a dual description where the relevant
degrees of freedom are getting more dilute, we can assume that it
also admits a coarse-grained description and simply ask the
question about how these two descriptions connect together. In
this way, all we need to do is to correctly and systematically
follow the (semi-)classical solutions of the theory based on the
truncation of string effective actions to graviton multiplet
two-derivative sector to near the string scale \cite{tseva}. In the string
frame the action is given by
\be S_e=\frac{1}{2 \kappa_N} \int d^{N+1}x \sqrt{-g_s} e^{-\phi_s}
\left(  R_s + (\partial \phi_s)^2 - {\cal L}_m \right) \, ,
\label{action} \ee
where ${\cal L}_m$ encodes the string ensemble contributions, the
effects of stringy corrections, and possibly any other sources of
energy density, including the dilaton potential. The dimensional
scale $\kappa_N = M_s^{2-N}$ is the string scale in $D=N+1$
dimensions, and $\phi_s$ is the string frame dilaton, related to
the string coupling by $g_s^2=e^{\langle \phi_s \rangle }$. The
subscript $s$ indicates a quantity computed in string conformal
frame, and not only a string gas source. For the exploration of
gravitational effects of coarse-grained string sources, it is
enough to explore the theory (\ref{action}) only on homogeneous
backgrounds. However, to elucidate the form of the string gas
source terms, following \cite{tseva} one can use the
anisotropic metrics
\be
ds^2=-n(t)^2 dt^2+\sum_{i=1}^N e^{2 \lambda_s^{(i)}(t)} \, dx_i^2 \, ,
\label{metric}
\ee
where $a^{(i)}_s(t) =e^{\lambda^{(i)}_s(t) }$ are the string frame
scale factors of the toroidal directions parameterized by $x^i$s,
and $n(t)$ is the lapse function, which can be gauge fixed to
unity. While such configurations are generically infested with
curvature singularities \cite{tseva}, where some dimensions
shrink to zero while others explode, away from the singularities
one can use thermodynamic arguments to deduce the form of the
string sources, and specifically energy density and pressures, at
large scales in the gravitational equations. So, specializing for
the moment to only ensembles of gravitating string gas,
dimensionally reducing the theory (\ref{action}) on the commuting
spatial Killing vectors $\partial_i$ of (\ref{metric}), and using
the shifted string frame dilaton $\varphi_s= \phi_s- \sum_{i=1}^N
\lambda_s^{(i)}$, one finds the effective $1D$ action
\cite{tseva},
\be S_{1D} = \int dt \, n \Bigg\{ \frac{e^{-\varphi_s}}{M_s}
\Bigl( \sum_i \frac{(\dot \lambda_s^{(i)})^2}{n^2} - \frac{\dot
\varphi_s^2}{n^2} \Bigr) + F(\lambda_s^{(i)}, n \beta) \Bigg\} \,
, \label{gasact} \ee
where $F$ is the comoving free energy of the string gas, which
depends on the temperature $\beta^{-1}$ and the scales of the
comoving `container' $\lambda_s^{(i)}$. Varying this action with
respect to intensive parameters and using the standard relations
between energy, pressure and free energy, $E_s = - (F + \beta
\partial_\beta F)$ and $P_i = \partial_{\lambda_s^{(i)}} F = -
(\partial_{\lambda_s^{(i)}} E)|_{\beta = const}$, in the gauge
$n=1$ one finds  the string frame Hamiltonian constraint and
equations of motion
\bea
\vd_s^2-\sum_{i=1}^N \left( \ld_s^{(i)} \right)^2
&=& e^{\varphi_s} E_s \, , \label{e1} \\
\ldd_s^{(i)}-\vd_s \ld_s^{(i)}
&=& \half e^{\varphi_s} P_s^{(i)} \, , \label{e2}\\
\vdd_s - \sum_{i=1}^N \left( \ld_s^{(i)}\right)^2
&=& \half e^{\varphi_s} E_s \, , \label{e3}
\eea
respectively, where we have chosen the units $M_s = 1$.  The
sources obey the conservation equation,
\be
\dot{E}_s+\sum_{i=1}^N \ld_s^{(i)} P_i=0 \, ,
\label{conseq}
\ee
which follows from (\ref{e1})-(\ref{e3}). We will focus on the
sources of the form $E_s= \sum_{i=1}^N E_{s0} \, e^{\beta
\lambda^{(i)}_s}$, in the limit when the distribution is
isotropic, $\lambda_s^{(i)} \equiv \lambda_s$. The pressures in
different directions $P_i$ are replaced by the mean pressure
$P_s=\frac{1}{N} \sum_{i=1}^N P_s^{(i)} = \frac{1}{N} \sum_{i=1}^N
\frac{\partial E_s}{\partial \lambda_s^{(i)}} =-\frac{\beta}{N}
E_s$, such that the equation of state parameters
$\gamma^{(i)}=P^{(i)}_s/E_s$ reduce in the isotropic case to
$\gamma=-\beta/N$.

Once the equations of motion (\ref{e1})-(\ref{e3}) are derived, it
is convenient to replace the comoving energy $E_s$ and comoving
pressure $P_s$ by the energy density $\rho_s$ and pressure $p_s$.
They are defined by $\rho_s = E_s e^{-N\lambda}$ and $p_s = P_s
e^{-N\lambda}$, such that the relevant sources in the equations of
motion (\ref{e1})-(\ref{e3}) can be substituted according to
\be e^{\varphi_s} E_s = e^{\phi_s} \rho_s \, , ~~~~~~~~~~~~~
e^{\varphi_s} P_s = e^{\phi_s} p_s \, . \label{enerpres} \ee
This allows one to rewrite the equations (\ref{e1})-(\ref{e3}) in
terms of the original $N+1$-dimensional dilaton variable $\phi_s$
as
\bea \left( \frac{N^2-N}{2} \right) \lds_s +\half \pds_s - N \ld_s
\pd_s &=&\half e^{\phi_s} \rho_s \, ,
\label{ee1}\\
\ldd_s- \pd_s \ld_s + N \lds_s &=& \half e^{\phi_s}  p_s \, ,\\
\dot{\rho}_s + N \ld_s \left( \rho_s + p_s \right) &=& 0 \, . \label{ee3}
\eea
The last equation is the stress-energy conservation, which follows
from (\ref{conseq}) and (\ref{enerpres}). The dilaton equation of
motion follows from stress-energy conservation, as one can verify
by taking a derivative of (\ref{ee1}) and using Eqs.
(\ref{ee1})-(\ref{ee3}) to solve for $\ddot \phi_s$. This form of
the equations of motion is convenient to go beyond the gravitating
string ensemble approximation and introduce more general sources
of stress-energy, which may depend on the dilaton field, including
any dilaton potential on the landscape. In that case, of course,
the simple equation of state $p_s = \gamma \rho_s$ will be
replaced by a more complicated expression, derived from the
Lagrangian. This will be useful to extend our results to more
general sources, particularly when formulating the conditions for
the connection between different gravitating string ensembles.

\subsection{Isotropic Single Fluid Cosmologies}

The equations of motion (\ref{e1})-(\ref{e3}) or
(\ref{ee1})-(\ref{ee3}) can be easily solved for the single fluid
component, with a constant equation of state $\gamma$. These
solutions serve as the attractors of the homogeneous and isotropic
multi-component fluid cosmologies. Indeed, starting with a
multi-component source, an expanding universe eventually ends up
being dominated by the component with the least positive equation
of state parameter $\gamma$. Conversely, the fluid with the most
positive $\gamma$ will asymptotically dominate in the contracting
universe limits. When the fluids are composed of unstable states
which may decay, the equation of state may eventually change,
transmogrifying the dominant source of stress-energy and the
universe it inhabits into a different fluid model. One may
therefore try to patch together general cosmologies from such
attractor geometries. The key for this quilt work is to determine
the rules for connecting solutions. At the leading order, one may
assume that the decay time is much faster than the Hubble time at
the moment of decay, and approximate the transition by an
instantaneous one. In this limit one can again ignore the
microscopics and retain the long range, coarse-grained
description, using Israel junction conditions to relate the field
variables at the spacelike boundary between different epochs. We
will pursue this approach in what follows.

The starting point is to taxonomize the various interesting single
component solutions. To count them up, one can first rewrite the
system of equations (\ref{ee1})-(\ref{ee3}) in the canonical form,
by solving the Hamiltonian constraint (\ref{ee1}). This yields
\bea
\dot{\phi}_s&=&N H_s \pm \sqrt{N H_s^2 
+ e^{\phi_s} \rho_s} \, , \label{br1}\\
\dot{H}_s&=&\frac{1}{2} e^{\phi_s}  p_s \pm H_s \sqrt{N H_s^2 
+e^{\phi_s} \rho_s}  \, , \label{br2}\\
\dot{\rho}_s&=&-N H_s \left( \rho_s + p_s \right) \, , \label{br3}
\eea
where we have introduced the notation $H_s \equiv \ld_s$. We
immediately note that the solutions can be classified by the sign
in front of the root, which by the definition of $\varphi_s =
\phi_s - N \lambda_s$ is precisely the sign of the reduced dilaton
flow, i.e. ${\tt sgn}(\dot \varphi_s)$, as we can verify by
comparing (\ref{e1}) and (\ref{br1}), and by the arrow of time.
The sign of $\dot \varphi_s$ can change only if the discriminant
$N \dot{\lambda}_s^2  + e^{\phi_s} \rho_s$ vanishes. Clearly, this
is impossible if $\rho_s \ge 0$. We will discuss this in more
detail later on. The sign of $H_s$ can change if the universe
`bounces', which may occur in the string frame due to the fact
that the dilaton kinetic term is not canonical. At any rate, this
discussion shows that for any value of the source terms there are
four solutions, described by the combinations of the two signs.
The latter sign selection is the same one encountered in the
pre-Big Bang cosmology, and so we adopt the same terminology here,
as it obviously generalizes. From now on, we will refer to the
$(\pm)$-branch by referring to the sign in front of the square
root.

Equations (\ref{br1})-(\ref{br2}) show that as long as $\rho_s \ge
0$, in the expanding universe on the $(+)$ branch, the string
coupling $g_s = e^{\langle\phi_s\rangle/2}$ increases. We also see
that on the $(-)$ branch, the string coupling increases during the
contracting phase. On the other hand, the evolution of the string
coupling in a $(+)$ branch contracting universe and a $(-)$ branch
expanding universe is more sensitive to the sources, and depends
on the time dependence of $e^{\phi_s} \rho_s$. In particular, it
may occur that the $\rho_s >0$ fluid can flip the sign of
$\dot \phi_s$ in these cases, as
can be seen by rewriting Eq. (\ref{br1}) as $\dot \phi_s = \pm
\sqrt{N} \sqrt{\dot \varphi_s^2 - e^{\phi_s} \rho_s} + \dot
\varphi_s$. This also shows that if $\rho_s<0$,
the coupling is fixed to grow on the $(+)$ branch contracting
solutions, and decrease on the $(-)$ branch expanding solutions.

We should note that we can generate new solutions by the symmetry
transformations of the effective action, which are the
time-reversal $dt \rightarrow -dt$ and $T$-duality \cite{gabr}. In fact, any
two solutions which reside on the {\it same} branch, as defined by
the sign of $\dot \varphi$, but differ by the sign of $H_s$, are
$T$-dual images of each other. The $T$-duality operates on the
background as the map $\lambda_s \rightarrow - \lambda_s$,
$\varphi_s \rightarrow \varphi_s$. This also implies the change of
sign of the equation of state parameter $\gamma = p/\rho$, which
is easy to see from the fact that the $T$-dual of the decaying
energy density increases, and vice-versa. An important thing to
stress is that $T$-duality does not change the branch since the
sign of $dt$, and therefore $\dot \varphi_s$, remains unchanged.
On the other hand, the time-reversal operation changes the branch
on which the solution resides, since under it the sign of $\dot
\varphi_s$ changes. Thus starting from any one solution, we can
get four different ones by the sequence of maps $T$-duality
$\rightarrow$ time-reversal $\rightarrow$ $T$-duality. In the case
$\gamma=0$, or when $p_s = \rho_s = 0$ as in pre-Big Bang
cosmology \cite{pbb}, this exhausts all the independent families
of solutions, but not in general cases, as we will see in what
follows.

Cosmic expansion and/or acceleration in the string frame do not
imply the same in the Einstein conformal frame, where the graviton
kinetic terms are normalized canonically. On the contrary, when
the dilaton evolves fast, the picture in the Einstein frame may be
completely reversed, so that an expanding string frame universe
looks like a collapsing universe crashing into a singularity. This
is relevant to understand the obstructions to pasting together
different cosmic epochs. To go to the Einstein frame, one performs
the conformal map to the action where the graviton and dilaton
terms are canonically normalized. The field redefinition which
relates the Einstein frame and the string frame variables is
\be g^{e}_{\mu \nu}=e^{-2 \phi_S/(N-1)} g^{s}_{\mu \nu} \, ,
\;\;\;\;\;\;\;\;\;\;\;\; \phi_e=\sqrt{\frac{2}{N-1}} \; \phi_s \,
, \label{conftrans} \ee
which on FRW backgrounds yields $a_e=e^{-\phi_s/(N-1)} a_s$.
Further, one must be careful with the comoving time relationship
as well, which is given by gauge-fixing the two lapse functions to
unity. This yields $d\tau_e = e^{-\phi_s/(N-1)} dt$. Using $\phi_s
= \varphi_s + N \lambda_s$, the logs of the scale factors and
their time derivatives are related by
\be \lambda_e=- \frac{\lambda_s + \varphi_s}{N-1}  \, ,
~~~~~~~~~~~ \frac{d\lambda_e}{d\tau_e} =-\left( \frac{\dot
\lambda_s + \dot \varphi_s}{N-1}\right) \frac{dt}{d\tau_e} \, .
\label{lambdas} \ee
From the second of (\ref{lambdas}) we see that the sign of the
Einstein frame Hubble parameter $H_e \equiv
\frac{d\lambda_e}{d\tau_e}$ can be reversed relative to the sign
of $H_s$, depending on the evolution of $\phi_s$. This can be seen
most easily on a case to case basis.

Let us now write down the exact single fluid isotropic solution,
defined by a constant equation of state parameter $\gamma =
P_s/E_s = p_s/\rho_s$ and the scaling law $E=E_0 \exp \left(
-\gamma N \lambda \right)$. For this purpose it is useful to go
back to the (isotropic limit of ) the equations
(\ref{e1})-(\ref{e3}) and introduce a new time coordinate $x$
defined by $E= \frac{dx}{dt}$. The equations of motion become
\bea \vps_s - N \lps_s&=& E_0^{-1}
e^{\varphi_s+\gamma N \lambda_s} \, , \\
\lpp_s-\gamma N \lps_s - \vp_s \lp_s &=&
\half \gamma E_0^{-1} e^{\varphi_s+\gamma N \lambda_s} \, , \\
\vpp_s-\gamma N \vp_s \lp_s -N \lps_s &=&
\half   E_0^{-1} e^{\varphi_s+\gamma N \lambda_s} \, ,
\label{dimsys}
\eea
respectively, where the prime denotes a derivative with respect to
$x$. The solutions of these equations are \cite{pbb,russ}
\bea
\lambda_s&=& \lambda_{s0}+ \frac{\gamma}{\alpha}
\ln \left[x(x-x_*) \right]+ \frac{1}{\alpha \sqrt{N}}
\ln \left( 1-\frac{x_*}{x} \right)\, , \label{1nlos} \\
\varphi_s&=&\varphi_{s0}-\frac{1}{\alpha} \ln \left[x(x-x_*) \right]
-\frac{\gamma \sqrt{N}}{\alpha} \ln \left( 1 -\frac{x_*}{x} \right)\, ,
\label{2nlos}\\
\phi_s&=&\phi_{s0}- \frac{1-N\gamma}{\alpha}\ln \left[x(x-x_*) \right]
- \frac{(\gamma-1)\sqrt{N}}{\alpha} \ln
\left( 1- \frac{x_*}{x} \right) \, , \label{3nlos}
\eea
where $\alpha=1-N\gamma^2$. The integration constant
$\lambda_{s0}$ is pure gauge, and can be changed arbitrarily by a
rescaling of the spatial coordinates $x^k \rightarrow \zeta x^k$.
The constant $\varphi_{s0}$ is linked to the comoving energy $E_0$
according to $e^{\varphi_{s0}} = \frac{4}{\alpha} \, E_0 \,
e^{-\gamma N \lambda_{s0}}$.  Note that in principle, we can
always write the solution in the form where the integration
constant $x_*$ is positive. Indeed, to do it, it is sufficient to
shift $x \rightarrow x + x_*$. In this case, the only change in
the solution (\ref{1nlos})-(\ref{3nlos}) is the sign flip of the
terms $\propto  \ln \left( 1- \frac{x_*}{x} \right)$. It is quite
clear that the times $x=0$ or $x=x_*$ are special moments in the
evolution of the epoch described by (\ref{1nlos})-(\ref{3nlos}),
as there the scale factor $a = e^{\lambda_s}$ can vanish or
diverge. To see precisely what goes on, however, it is more
convenient to study separately the cosmologies of different
relevant string phases.

\subsection{Hagedorn Phase}

In string theory, the density of states diverges at some high
temperature, where the theory undergoes Hagedorn transition.
Beyond this scale, the injection of additional energy into the
system does not affect the temperature, which remains constant,
but simply increases the entropy \cite{hagedorn}. Essentially, what happens is
that the strings become extremely wiggly as they keep absorbing
the energy \cite{atwitt}. This saturation of the temperature implies that the
dependence on the other intensive parameters vanes, and so the
pressures vanish. Hence the Hagedorn phase is described by the
constant comoving energy $E=E_0$ and vanishing pressure $P=0$,
giving the equation of state parameter $\gamma=0$ \cite{tseva}. This implies
that the parameter $\alpha$ as defined in the solution
(\ref{1nlos})-(\ref{3nlos}) is $\alpha=1$. Rewriting this solution
in terms of the comoving string frame time coordinate, by using $t
= x/E_0$, and setting $x_*/E_0 = t_*$, we find \cite{tseva}
\bea
\lambda_s&=&\lambda_{s0}+\frac{1}{\sqrt{N}}
\ln\left( 1-\frac{t_*}{t} \right) \, , \nonumber \\
\varphi_s&=&\varphi_{s0}- \ln\left[  t \left(t-t_* \right) \right] \, .
\label{sol2}
\eea
The unshifted dilaton is
\be
\phi_s=\phi_{s0}-2\ln |t|+\left( \sqrt{N}-1 \right)
\ln \left(  1-\frac{t_*}{t} \right) \, ,
\label{hagdil}
\ee
so that the string coupling is $g_s = \frac{g_{s0}}{|t|}\left(
1-\frac{t_*}{t} \right)^{\frac{\sqrt{N}-1}{2}}$.

First we note immediately that the special solution with $t_*=0$
describes a static string geometry, with $\lambda_2 = \lambda_{s0}
= {\rm const}$ and $\varphi_s = \varphi_{s0} - 2\ln t$, where the
runaway dilaton absorbs the effects of the Hagedorn source
allowing the universe to loiter forever. This special solution is
{\it not} a universal attractor, since it is unstable: while there
are configurations which flow towards it, representing either
universes expanding from zero size or contracting from infinite
size (as exemplified by the Class III and IV solutions described
below), their time-reversals, which are perfectly allowed by the
time-reversal symmetry of the field equations, describe the
evolution away from this special geometry, either towards a crunch
or an infinite universe.

Now, the solutions for $t_* \ne 0$
have classical curvature singularities at $t=0$ or $t = t_*$,
where the scale factor either diverges or vanishes. The interval
between $0$ and $t_*$ is classically forbidden, since inside it
the logarithms are imaginary. Note that the ordering $0$ and $t_*$
does not introduce extra solutions because the order can be changed
by the shift $t \rightarrow t +
t_*$, which sets $t_* >0$ while flipping the sign of the log term
in the equation for the scale factor. This means, that the
solutions with $t_*<0$ are simply $T$-duals of the solutions with
$t_*>0$ for a fixed sign of $t$. On the other hand, the
time-reversed geometries are obtained by
simultaneously flipping the signs of both $t$ {\it and} $t_*$ in
the solution. The solutions (\ref{sol2}) and (\ref{hagdil}) are
perturbatively meaningful descriptions of the following classical
string geometries:

\bi
\item
\noindent {\bf Class I:} The universe expands from finite size,
starting with $g_s \ll 1$, on the interval $t\le 0$, with $t_* >0$
and $\dot \varphi_s >0$ (evolving away from the $t_*=0$ repeller)
until $t=0$, where the scale factor and the string coupling
diverge, and yield a curvature divergence as well. Effective field
theory description will cease in the course of evolution towards
this limit.

\item \noindent {\bf Class II:} $T$-dual of Class I,
which contracts on the interval $t\le 0$, with $t_*>0$ and $\dot
\varphi_s >0$ (evolving away from the $t_*=0$ repeller) from
finite size to a Big Crunch at $t=0$, while the dilaton flows to
weak coupling $g_s \ll 1$. Geometric description ceases in this
limit.

\item \noindent {\bf Class III:} The universe slowly
expands for $t \ge t_*>0$ from zero size to a constant size at $t
\rightarrow \infty$, with $\dot \varphi_s<0$ (approaching the $t_*
= 0$ attractor) and the dilaton grows, $\pd_s>0$, such that the
string coupling becomes strong at late times, and weakly coupled
perturbative picture must eventually be replaced by a strongly
coupled description. During the early epoch when $a <1$, the
geometric description fails. This solution is in fact
time-reversal of the Class II solution.

\item \noindent {\bf Class IV:} $T$-dual of Class III,
which during times $t \ge t_*>0$ contracts from infinite size to a
constant finite size at $t\rightarrow \infty$ with $\dot
\varphi_s<0$  (approaching the $t_*=0$ attractor) and has a finite
string coupling, that initially grows ($\pd_s>0$) but eventually
decreases ($\pd_s<0$), evolving to weak coupling regime. Again,
the geometric description of this solution cannot be trusted
beyond $a <1$. This solution is time-reversal of the Class I
solution.

\ei

\begin{figure}[]
\centerline{\hbox{ \hspace{0.0in}
\epsfxsize=5in \epsffile{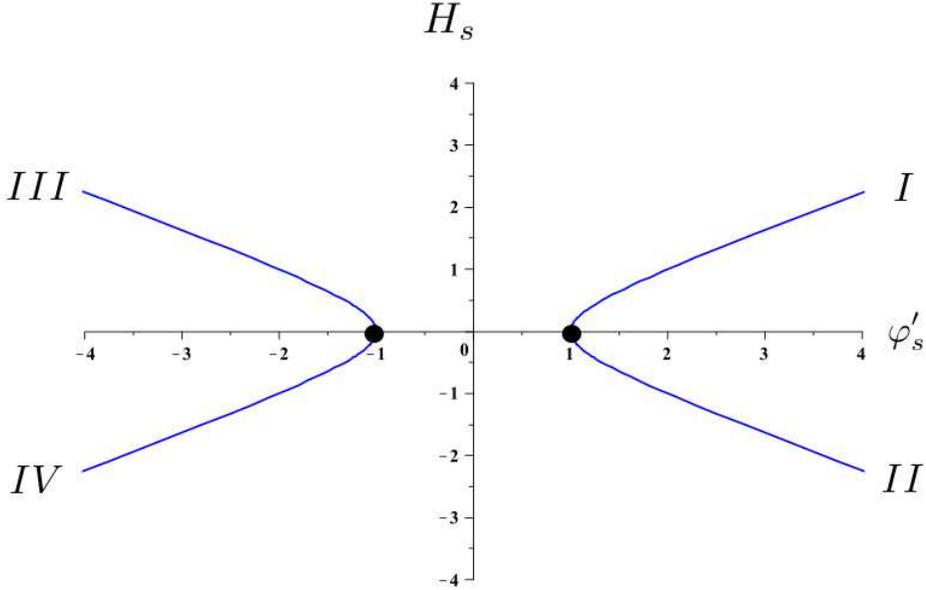} \hspace{0.25in}}}
\caption{\label{fig1} The  $(\vp_s, H_s)$ phase diagram of
Hagedorn cosmologies. The black dots mark the Hagedorn fixed
points, $(\vp_s, H_s )=(\pm \sqrt{E_0},0)$.}
\end{figure}

As explained above, there us no more independent solutions,
essentially because $\lambda_s$ depends only on the ratio
$\frac{t-t_*}{t}$. This degeneracy follows from $\gamma=0$, so
that all the gravitating Hagedorn phases remain connected by
duality and time-reversal. The solutions are summarized in Table
(1). Note that the $(+)$ branch solutions evolve towards the
singularity, and $(-)$ branch away from it. The evolution of these
solutions is represented in the $\dot \varphi_s, H_s$ phase plane
in Figure (\ref{fig1}), with the convention that the time flows
from left to right.

\vskip.5cm
\begin{table}[htdp] \begin{center}\begin{tabular}{|c|c|c|c|c|c|c|c|}
\hline  Class &  Branch &\;\;\; Expansion \;\;\; & \;\;\;
Shifted Dilaton \;\;\; & \;\; Time \;\; &  \;\; Singularity \;\;  \\
\hline
I & $(+)$ &$H_s>0$ & $\vpd_s>0$ & $t<0$ & Future \\
II & $(+)$ & $H_s<0$ & $\vpd_s>0$ &  $t<0$  & Future \\
III & $(-)$ & $H_s>0$ & $\vpd_s<0$ & $t>t_*$ & Past \\
IV & $(-)$  & $H_s<0$ & $\vpd_s<0$ &  $t>t_*$ & Past \\
\hline \end{tabular} \caption{The four Classes of the Hagedorn
solution (\ref{sol2}). In Class II, if $(\sqrt{N}+1)/2>t/t_* >1 $
then $\pd_s>0$, otherwise $\pd_s<0$.  Likewise, in Class III if
$(\sqrt{N}+1)/2 < t/t_*< 1$ then $\pd_s<0$, otherwise $\pd_s>0$.
Class I and II, and Class II and IV, respectively, are related by
$T$-duality, and Class I and IV and II and III, respectively, by
time-reversal. }
\end{center}
\label{table1}
\end{table}
\vskip-.5cm

The perturbative description is limited by the flow of a solution
to short distances and/or strong coupling. The geometrical
singularities appear particularly dangerous. Of course, once we
recall the philosophy behind dualities, the singularities may be
interpreted as merely an indicator of dual branches which the
solution must be replaced by, once the scale factor dives below
the string scale. The question then is if such dual maps relate
geometries that admit asymptotic description in some supergravity.
In particular, one would like to be able to have a cosmology whose
late time behavior describes a radiation dominated stage, with the
coupling possibly evolving away from extreme weak coupling, so
that the theory may yield useful mechanisms for moduli
stabilization and asymptote to a phenomenologically reasonable
regime. To see if such a stage can be preceded by a Hagedorn
stage, we must first explore the phase space of cosmology of
excited dilute strings, to which we turn to next.

\subsection{Dilute String Phase}

Below the Hagedorn temperature, the ensemble of strings behaves
differently. If spatial directions are compact, then the dilute
strings can carry energy in two different channels: i) the
momentum modes, which describe the propagation of disturbances
along the string, and ii) the winding modes, which store the
energy to stretch the string around the compact dimension. In the
former case, the momentum mode channels are relativistic at high
energies, and so their thermal ensemble behaves as a gas of
radiation in the string frame, with the equation of state
parameter obeying $\gamma=1/N$, yielding $E=E_0 e^{-\lambda}$. The
winding modes are the $T$-dual of the momentum modes, and so their
equation of state parameter, by the fact that under $T$-duality
the pressure changes sign, is $\gamma=-1/N$, such that $E=E_0
e^\lambda$ \cite{gabr,tseva}. The negative pressure reflects the fact that when the
dimensions expand, the winding mode energy {\it increases} since
it is proportional to the length of the string. Hence the energy
density dilutes more slowly: although the number density drops
down as is usual, with the one power of spatial volume, the energy
increases and compensates some of the dilution. Of course, in the
case when strings gravitate, the whole geometry needs to be
consistently subjected to the $T$-duality transformation.

The homogeneous cosmology dominated by momentum or winding modes
is still given by the general solution (\ref{1nlos})-(\ref{3nlos})
but with the parameter $\alpha$ now given by
$\alpha=\frac{N-1}{N}$. Thus the solution for the scale factor and
the shifted dilaton formally reads
\bea \lambda_s&=& \lambda_{s0} \pm  \frac{1}{ N-1}
\ln \left[ x(x-x_*) \right] + \frac{\sqrt{N}}{N-1}
\ln \left( 1-\frac{x_*}{x} \right) \, , \label{radsoln1}\\
\varphi_s&=&\varphi_{s0} -\frac{N}{N-1} \ln \left[ x(x-x_*)
\right] \mp \frac{\sqrt{N}}{N-1}  \ln \left( 1-\frac{x_*}{x}
\right) \, , \label{radsoln2} \eea
where the upper sign formally describes a momentum mode dominated
universe, and the lower sign a winding mode dominated universe.
The singular points of the solutions are as before $0$ and $x_*$,
and the interval between them is classically forbidden, by the
condition of reality of the solution. The unshifted dilaton in a
universe dominated by the momentum modes, which evolve as string
radiation, is then given by
\be \phi^{m}_s=  \phi_{s0}+\sqrt{N} \ln \left( 1-\frac{x_*}{x}
\right) \, , \label{radsoln3m} \ee
whereas in the universe dominated by the winding modes it is
\be
\phi^{w}_s=  \phi_{s0} - \frac{4N}{N-1} \ln |x| +
\sqrt{N} \frac{\sqrt{N}-1}{\sqrt{N}+1}
\ln \left( 1-\frac{x_*}{x} \right) \, , \label{radsoln3w}
\ee
as dictated by $T$-duality.

To see that the solutions with the opposite signs of $\gamma$ are
in fact $T$-duals of each other, we need to shift the `time' by $x
\rightarrow x+x_*$ and flip the sign of the integration constant
$x_*$. With this, it becomes clear that $\lambda_s - \lambda_{s0}
\rightarrow - (\lambda_s - \lambda_{s0})$ while $\varphi_s$
remains unaffected. In fact, this becomes obvious if we write the
expressions for the scale factor and the exponent of the shifted
dilaton,
\bea a_s &=& e^{\lambda_{s0}}
x^{-\frac{\sqrt{N}\mp1}{(\sqrt{N}-1)(\sqrt{N}+1)}}
(x-x_*)^{\frac{\sqrt{N}\pm1}{(\sqrt{N}-1)(\sqrt{N}+1)}} \, , \nonumber \\
e^{\varphi_s/2} &=&
\frac{e^{\varphi_{s0}/2}}{x^{\sqrt{N}\frac{\sqrt{N}\pm
1}{(\sqrt{N}-1)(\sqrt{N}+1)}}
(x-x_*)^{\sqrt{N}\frac{\sqrt{N}\mp1}{(\sqrt{N}-1)(\sqrt{N}+1)}}}
\, . \label{dualsolns} \eea
However, unlike before, the solutions depend on different powers
of $x$ and $x-x_*$, so that the sequence of $T$-duality and
time-reversal cannot link all of them directly. Instead, there are
two different families of linked solutions, which arise because
$T$-duality interchanges momentum and winding modes, and flips the
sign of $\gamma \ne 0$, breaking the degeneracy found in the
gravitating Hagedorn gas example.

To understand the universes which these solutions describe, we
need to follow their evolution for specific values of $x_*$. As
before, there are the special solutions with $x_*=0$, which are
\be \lambda_s = \lambda_{s0} \pm  \frac{2}{ N-1} \ln |x| \, ,
~~~~~~ \varphi_s= \varphi_{s0} -\frac{2N}{N-1} \ln |x|  \, ,
~~~~~~ \phi_s=  \phi_{s0} + \frac{2N(1\pm1)}{N-1} \ln |x| \, .
\label{radsolnspec} \ee
They are clearly $T$-duals of each other, representing asymptotic
attractors for momentum mode-dominated and winding mode-dominate
cosmologies, respectively. For general $x_* \ne 0$, there are
eight distinct solutions that comprise two families interlinked by
$T$-duality and time reversal, which are are conveniently
parameterized by the branch sign and the signs of $\gamma$ and
$x_*$. They are as follows:

\bi
\item
\noindent {\bf Class 1:} The momentum mode-dominated universe
which expands forever as a power law, starting with $g_s \ll 1$,
on the interval $x\ge x_* >0$, with $\dot \varphi_s<0$, such that
asymptotically $g_s \rightarrow g_{s0} = {\rm const}.$, evolving
towards the $x_*=0$ momentum mode attractor. This is a $(-)$
branch solution. Near the singularity $x=x_*$ the geometric
description will cease, and the classical geometry there resembles
a Big Bang.

\item \noindent {\bf Class 1$^{\tt TR}$:}
The time-reversal of the Class 1 solution, found by flipping the
signs of both $x$ and $x_*$. Evolves away from the $x_*=0$
momentum mode repeller.

\item \noindent {\bf Class 2:}
Another momentum mode-dominated universe, which however resides on
the interval $x<0$, with $x_*>0$, and which starts out as a
collapsing universe but undergoes a string frame bounce, and runs
off to a super-inflating pole singularity at $x=0$. This is a
$(+)$ branch solution, $\dot \varphi_s >0$, starting with some
$g_{s0} = {\rm const}$, evolving away from the $x_*=0$ momentum
mode repeller towards strong coupling $g_s \gg 1$. Near the pole,
both the geometric description and perturbation theory in $g_s$
break down. This solution cannot be related to the Class 1 case
neither by $T$-duality nor time-reversal, which is clearly from
the existence of the string frame bounce.

\item \noindent {\bf Class 2$^{\tt TR}$:}
The time-reversal of the Class 2 solution. Again, the signs of
both $x$ and $x_*$ are flipped. Evolves towards the $x_*=0$
momentum mode attractor.

\item \noindent {\bf Class 3:}
$T$-dual of the Class 1 solution, describing a winding
mode-dominated universe, which is collapsing from infinite size at
$x=x_* >0$ to zero as $x \rightarrow \infty$, with $\dot \varphi_s
<0$. This is a $(-)$ branch solution. In the course of evolution,
the string coupling changes from $g_s \gg 1$ to $g_s \ll 1$.
Initially, the theory is given by strongly coupled winding mode
gas, which evolves towards weakly coupled dense gas, described by
the $x_*=0$ winding mode attractor, where the geometric picture
breaks down.

\item \noindent {\bf Class 3$^{\tt TR}$:}
The time-reversal of the Class 3 solution. Evolves away from the
$x_*=0$ winding mode repeller.

\item \noindent {\bf Class 4:}
$T$-dual of Class 2, a winding mode-dominated universe defined on
the domain $x<0$ (with $x_*>0$) which initially slowly expands out
of zero size, evolving away from the $x_*=0$ winding mode
repeller. It reaches a maximum size due to the resistance from the
winding modes, and starts to collapse again. In the course of
evolution, $\dot \varphi_s >0$ so this is a $(+)$ branch solution.
The string coupling varies from $g_s \ll 1$ to $g_s \gg 1$, so
that near the crunch both the geometric description and $g_s$
expansion fail.

\item \noindent {\bf Class 4$^{\tt TR}$:}
The time-reversal of the Class 4 solution. Evolves towards the
$x_*=0$ winding mode attractor.

\ei

Rewriting these solutions in terms of the comoving time $t$ is
cumbersome. However in the asymptotic limits $|x| \rightarrow \pm
\infty$ it is straightforward to extract the leading order form of
the solution as a function of $t$. In fact these limits are
exactly the attractors (\ref{radsolnspec}). Using $dx/dt = E_0
e^{-\lambda_s}$, and the fact that (\ref{radsoln1}) converges to
$\lambda_s \rightarrow \pm \frac{1}{N-1} \ln x^2$ when $|x|
\rightarrow \infty$, we find that $|x| \sim |t|^{(N-1)/(N+1)}$
when momentum modes dominate. Similarly, when winding modes
dominate, $|x| \sim |t|^{(N-1)/(N-3)}$ for $N\ne 3$, and $|x| \sim
e^{E_0 |t|}$ for $N=3$, as is straightforward to see. Therefore
the momentum mode solutions in the limit $x \rightarrow \pm
\infty$ approach
\be a_s \to a_{s0} |t|^{\frac{2}{N+1}} \, , ~~~~~~~~ \varphi_s \to
\varphi_{s0} -\frac{2N}{N+1} \ln |t| \, , ~~~~~~~~ \phi_s \to
\phi_{s0} \, , \label{momasympt} \ee
which is precisely the form of an FRW radiation-dominated
cosmology in $N+1$ dimensions. The fact that the string dilaton
$\phi_s$ approaches a constant in this limit despite the absence
of a stabilizing potential is a well known consequence of the
conformal symmetry of the sources which has this attractor
behavior \cite{dilatons}. On the other hand, as $x \to \pm \infty$
the winding mode solutions converge to
\be a_s \to \frac{a_{s0}}{|t|^{\frac{2}{N-3}}} \, , ~~~~~~~~
\varphi_s \to \varphi_{s0} -\frac{2N}{N-3} \ln |t| \, , ~~~~~~~~
\phi_s \to \phi_{s0} - \frac{4N}{N-3}  \ln |t| \, ,
\label{winasympt} \ee
for $N \ne 3$, which changes to
\be a_s \to {a_{s0}} e^{-E_0 |t|} \, , ~~~~~~~~ \varphi_s \to
\varphi_{s0} - 3 E_0 \ln |t| \, , ~~~~~~~~ \phi_s \to \phi_{s0} -
6 E_0  |t| \, , \label{winasymptn3} \ee
when $N=3$. We therefore see that the winding mode geometries
describe an inflating or deflating asymptotic cosmology in the
string frame, depending on the direction of the flow of $t$.

\begin{figure}[htpd]
\centerline{\hbox{ \hspace{0.0in}
\epsfxsize=5in \epsffile{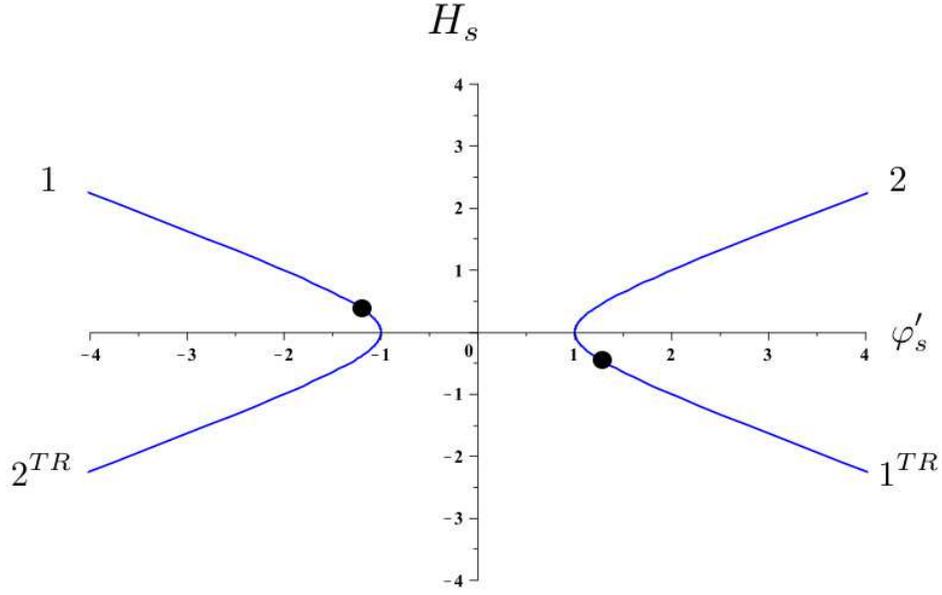} \hspace{0.25in} }}
\caption{\label{fig2}  The  $(\vp_s, H_s)$ phase diagram of
momentum mode-dominated cosmologies. The black dots mark the
momentum mode fixed point cosmologies, which in the $\tau$
variable are $ (\vp_s, H_s)=(\mp \sqrt{3E_0/2}, \pm
\sqrt{E_0/6})$. }
\end{figure}
\begin{figure}[htpd]
\centerline{\hbox{ \hspace{0.0in}
\epsfxsize=5in \epsffile{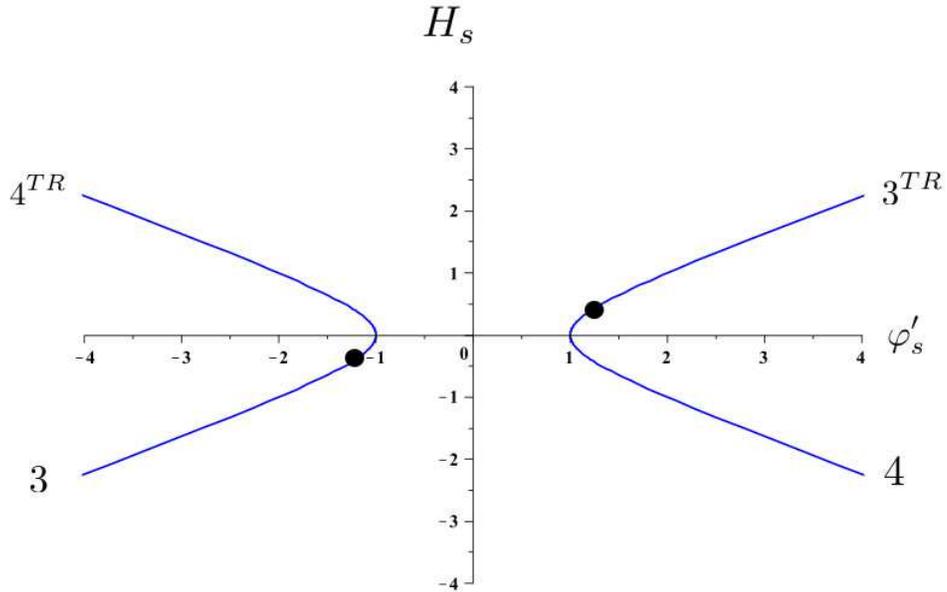} \hspace{0.25in} } }
\caption{\label{fig3} The  $(\vp_s, H_s)$ phase diagram of winding
mode-dominated cosmologies. The black dots mark the winding mode
fixed points, $ (\vp_s, H_s)=(\pm \sqrt{3E_0/2}), \pm
\sqrt{E_0/6})$. }
\end{figure}

These solutions are depicted in Figures (\ref{fig2}) (momentum
mode-dominated cosmologies) and (\ref{fig3}) (winding
mode-dominated cosmologies), where we use the time variable
$\tau$, defined by the relation $d\tau = \exp({\frac{\varphi_s -
\gamma N \lambda_s}{2}}) \, dt$ (see below), and assume that the
time flows again from left to right. The two families of solutions
mutually linked by time-reversal and $T$-duality are also
summarized in Tables (2) and (3). Again, we see that the $(+)$
branch solutions run to a singularity, while the $(-)$ branch ones
evolve away from it. We can now turn to determining how and when
these solutions may be matched onto the Hagedorn gas cosmologies
of the previous section.
\vskip.5cm
\begin{table}[htdp]
\begin{center}\begin{tabular}{|c|c|c|c|c|c|c|c|}
\hline  Class &  Branch & \;\;\; Expansion \;\;\; & \;\;\;
Shifted Dilaton \;\;\; & \;\; Time \;\;  &  \;\; Singularity \;\;  \\
\hline
1 \,\, & $(-)$ & $H_s>0$ & $\vpd_s<0$ & $x\ge x_*>0$ & Past \\
$1^{\tt TR}$ & $(+)$ & $H_s<0$ & $\vpd_s>0$ & $ x \le x_*<0$ & Future  \\
3 \,\, & $(-)$ &$H_s<0$ & $\vpd_s<0$ &  $ x \ge x_*>0$ & Past \\
$3^{\tt TR}$ & $(+)$ &  $H_s>0$ & $\vpd_s>0$ & $ x \le x_*<0$ & Future   \\
\hline \end{tabular} \caption{ Smooth family, without string frame
bounce. Classes 1 and 3 are $T$-duals of each other, as are their
time-reversed solutions, describing momentum mode-dominated and
winding mode-dominated cosmologies, respectively. }
\end{center}
\label{table2}
\end{table}
\vskip-.5cm
\begin{table}[htdp]
\begin{center}\begin{tabular}{|c|c|c|c|c|c|c|c|}
\hline  Class &  Branch & \;\;\;\; Expansion \;\;\;\; & \;\;
Shifted Dilaton \;\; & \;\; Time \;\;  & \, Singularity \, \\
\hline
2 \,\, & $(+)$ & $H_s<0 \to H_s >0 $ & $\vpd_s>0$ & $x<0$, $x_*>0$ & Future \\
$2^{\tt TR}$ & $(-)$ & $H_s > 0 \to H_s<0$ & $\vpd_s<0$ & $x >0$, $x_*<0$ & Past \\
4 \,\, & $(+)$ &$H_s > 0 \to H_s<0$ & $\vpd_s>0$ &  $ x <0$, $x_*>0$ & Future \\
$4^{\tt TR}$ & $(-)$ &  $H_s < 0 \to H_s>0$ & $\vpd_s<0$ & $ x>0$, $x_*<0$  & Past \\
\hline \end{tabular} \caption{ A family with the string frame
bounce. Classes 2 and 4 are $T$-duals of each other, as are their
time-reversed solutions, again describing momentum mode-dominated
and winding mode-dominated cosmologies, respectively.}
\end{center}
\label{table3}
\end{table}

\section{Branch Changes and Energy Conditions}

The transitions between different phases of string are normally
precipitated by the change in internal energy. As the dilute
string gas is heated to the Hagedorn temperature, it undergoes the
Hagedorn transition, and starts to absorb energy by increasing its
entropy while keeping a constant temperature. Conversely, if the
internal energy is dissipated, by decreasing entropy, the Hagedorn
gas will reach the transition point and begin to cool. An obvious
arena for such transitions is an expanding universe, where the
expansion dials the internal energy of the matter up and down.
Thus in string theory, it is expected that in cosmological
backgrounds the gas of string will scan through various allowed
thermodynamic phases as the universe evolves. This has
led \cite{brava,tseva} to propose models of early universe which
may even provide a way around the cosmological singularity, not
directly by a bounce but rather by a dual relationship between
collapsing and expanding universes. In these works, typically one
seeks to have cosmologies which look like a patchwork of different
thermodynamic phases of the string, that evolve into each other
dynamically.

In the presence of string gravity, however, the transitions
between different string phases become subtler, complicated by the
presence of the dilaton. The dilaton may significantly affect the
evolution of the background geometry inhabited by the string gas.
If one treats it too cavalierly, one may not notice that the
dilaton can obstruct transitions between some of the gravitating
thermodynamic string ensembles. One may argue that just before
such transitions the dilaton is stabilized by dynamical effects \cite{break,ali}.
But if the theory is to retain its duality symmetries at the
fundamental level, the dilaton stabilization and decoupling should
only occur at low energies, below which the dualities are obscured
by the infra-red effects. Hence one expects that any phase
transitions at high energies, in a very early universe, should be
subjected to the selection rules dictated by the full, $T$-duality
symmetric dilaton gravity which governs the geometry. Let us now
determine these rules in a precise manner. We will show that as
long as the energy density inhabiting the universe is
non-negative, branch changing is impossible, and the transitions
can only occur between string phases on the same geometric branch.
Further, if positivity of energy is not maintained, branches may
change, but unless the sources also violate 
NEC, which in homogeneous and isotropic cosmologies requires $p
+ \rho < 0$, it is impossible to go from a $(+)$ branch solution
to a $(-)$ branch one. Since $(+)$ solutions have future
singularities, this result is really a singularity theorem in a
different guise: it implies that the solutions cannot be relieved
of strong curvatures unless NEC is violated.

Indeed, when discussing the equations (\ref{br1})-(\ref{br3}), we
already observed the first part of this statement, that the
solutions on opposite branches, as characterized by ${\tt
sgn}(\dot \varphi)$, cannot link up unless the discriminant $N
\dot \lambda_s^2 + e^{\phi_s} \rho_s$ vanishes. Since the first
term is positive definite, this implies that if the energy density
$\rho_s$ is also positive definite, a transition between different
branches is {\it impossible}. Hence to have any chance for
changing branches, the energy density must dip below zero, which
of course by itself still does not guarantee that the branch
change will actually occur.

Although very simple, this statement of positivity of energy
excluding branch changing is already quite strong: it already
excludes some of the phase transitions described in the
literature. For example, \cite{tseva} describe a universe which
expands away from the self-dual point on the Hagedorn phase, and
eventually below the Hagedorn transition it transits to a
radiation phase, or possibly to an expanding winding mode phase.
Near the self-dual point, the universe just hovers, with the
background asymptotically evolving out of the attractor solution,
given by the $t_* = 0$ limit of (\ref{sol2}). Prior to the
self-dual point, the universe may have evolved as a time-reversed
cosmology of either of the two post self-dual point options.
However, the inspection of the Hagedorn phase solutions, in Table
(1), immediately shows that the Hagedorn cosmology that slowly
flows out of the fixed point at the self-dual radius is the Class
I geometry, which is a $(+)$ branch solution. On the other hand,
the radiation cosmology which has no future singularities but
evolves towards a conventional FRW universe with a frozen dilaton
is the Class 1 geometry, a $(-)$ branch solution, as we see from
Table (2). Thus if the energy density dominating the universe is
always positive definite, the decay of the Hagedorn gas in a
cosmology evolving from the attractor {\it will not} yield an
asymptotic radiation-dominated FRW universe. Alternatively, the
Hagedorn gas stage is really evolving towards the strong curvature
regime, whereas the radiation FRW universe is not. To change from
one to another, the evolution to large curvature must be averted,
and that is impossible with the positive definite energy density.

If we turn to winding mode cosmologies, we can see by a similar
argument as above that the Class I Hagedorn gas universe, which
flows out of the attractor, can only connect to the Class 4
geometry, which is also a $(+)$ branch solution (see Table (3)).
This transition is possible early on while the winding mode
universe is still expanding. However, the expansion ceases in this
universe and it collapses back to a future singularity. Once it
passes below the self-dual radius, we may replace it by its
$T$-dual description, however this is the Class 2 momentum-mode
dominated cosmology which experiences a pole singularity in the
future as opposed to asymptoting to the conventional
radiation-dominated FRW universe. The bottomline is that the
strong curvature singularity in the future is unavoidable. Similar
obstructions can be inferred for the time-reversed configurations
describing the universe before the self-dual point.

Thus as long as the energy density is non-negative, we can only
sew together the solutions on the {\it same} branch \cite{kalmad}. For instance,
requiring that the late universe is a radiation-dominated FRW
universe of Class 1 confines one to the $(-)$ branch solutions. If
the evolution is then followed backward in time, it is plausible
that this universe can be preceded by the Hagedorn cosmology given
by the Class III solution, which is on the $(-)$ branch too.
However, in the far past, this solution does {\it not} asymptote
to the Hagedorn attractor configuration, but instead has a Big Big
singularity in the string frame. It is conceivable that the small
universe region near this singularity can be resolved as the
$T$-dual geometry of Class IV, which in turn could have evolved
from the Class 3 collapsing winding mode-dominated world. In this
chain of events, one might be able to `relegate' the problem of
cosmological singularity, reinterpreting it as the problem of
finding the correct dual description of a small universe. On the
other hand, in contrast to the discussion of the cosmic phase
quilt that was discussed in \cite{tseva}, the universe of this
example does {\it not} spend an infinite time lingering around the
Hagedorn attractor. In fact, the Hagedorn attractors are
completely excised, now being in the future of the transition to
FRW radiation, or in the past of the winding mode condensation.
One could now imagine that the local perturbations in the
collapsing $T$-dual geometry, if such is consistently constructed,
can be computed, and that they could map onto long distance
perturbations in the small universe near the singularity. This
would resolve their short distance dynamics. However this would be
completely different from the ideas for generation of
perturbations in string cosmology which have been pursued so far.
Nevertheless any applications of such ideas to phenomenology must
await the development of the framework that might enable us to
compute how the local perturbations in one universe $T$-dualize to
large scale perturbations in another. The present techniques only
allow a precise construction of $T$-dual geometries when there are
isometries, which are of course all generically broken by the
perturbations.

\begin{figure}[htpd]
\centerline{\hbox{ \hspace{0.0in}
\epsfxsize=4.5in \epsffile{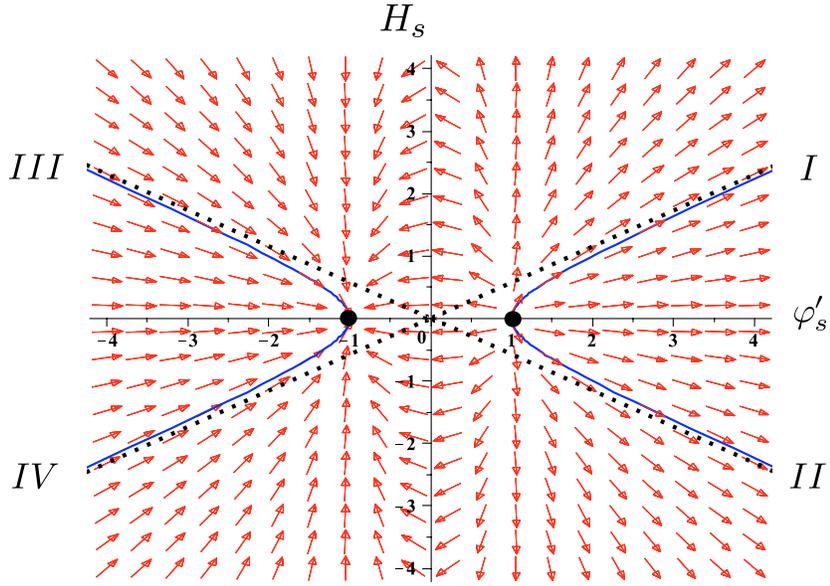} \hspace{0.25in} } }
\caption{\label{fig4} The $(\vp_s, H_s)$ phase diagram of Hagedorn
cosmologies, with phase space flow and the limiting envelopes
(dotted lines) describing the case $E_0=0$. }
\end{figure}

\begin{figure}[htpd]
\centerline{\hbox{ \hspace{0.0in}
\epsfxsize=4.5in \epsffile{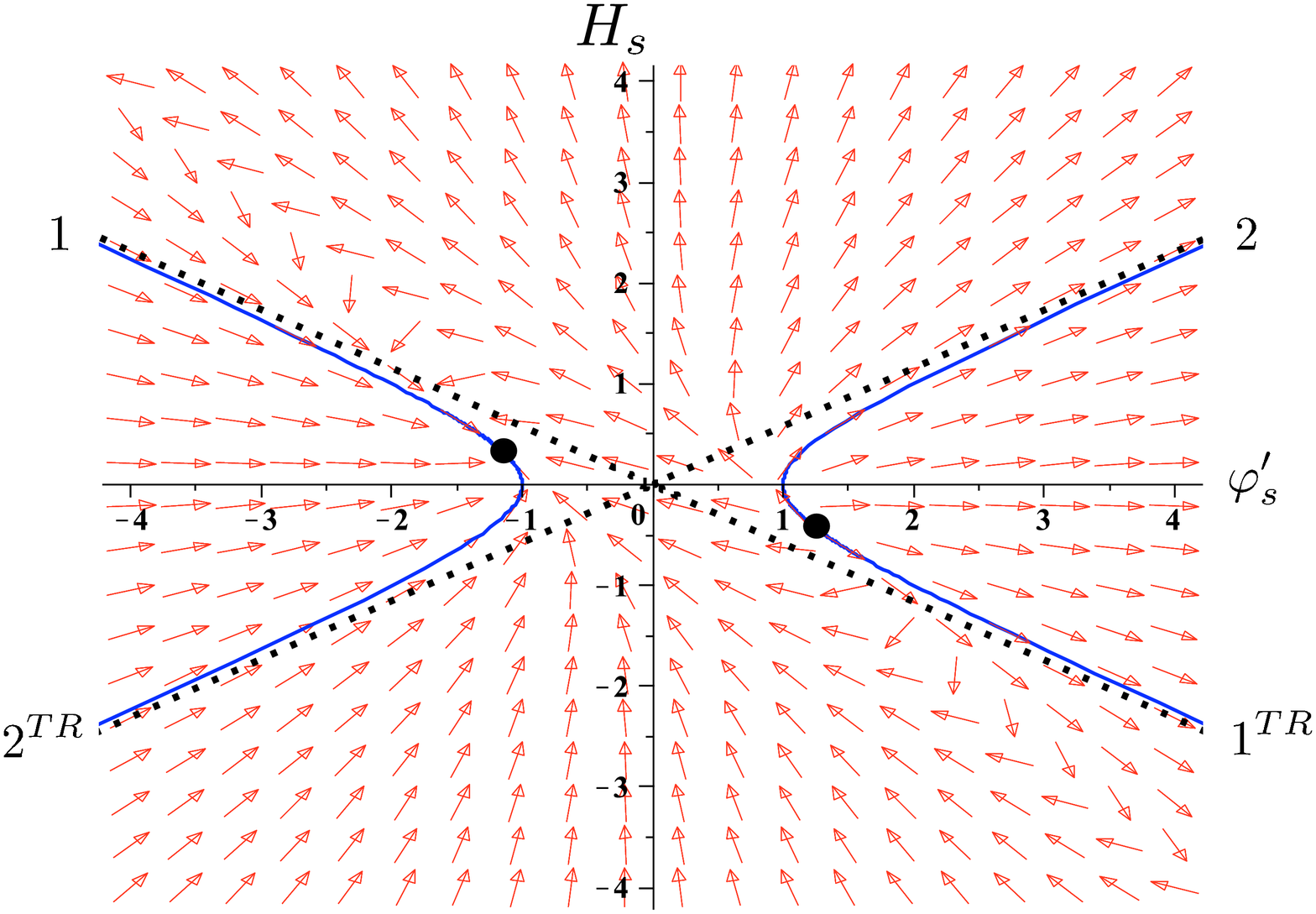} \hspace{0.25in} } }
\caption{\label{fig5} The $(\vp_s, H_s)$ phase diagram of momentum
mode cosmologies, with phase space flow and the limiting envelopes
(dotted lines) describing the case $E_0=0$. }
\end{figure}

\begin{figure}[htpd]
\centerline{\hbox{ \hspace{0.0in}
\epsfxsize=4.5in \epsffile{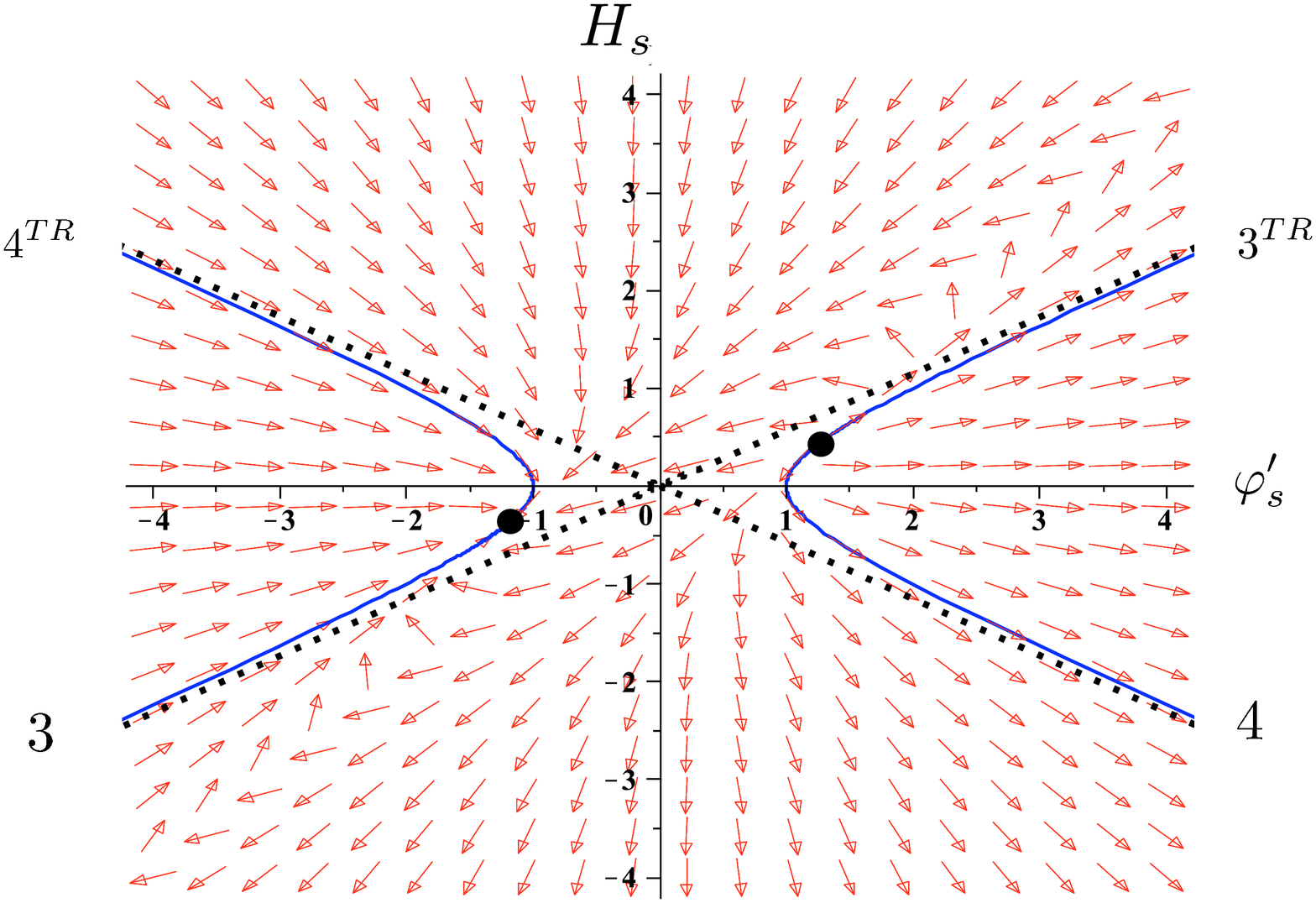} \hspace{0.25in} } }
\caption{\label{fig6} The $(\vp_s, H_s)$ phase diagram of winding
mode cosmologies, with phase space flow and the limiting envelopes
(dotted lines) describing the case $E_0=0$. }
\end{figure}

To understand the dynamics of the solutions with non-negative
energy density, it is convenient to go to the phase space
description. The phase space evolution of solutions can be
illustrated in simple cases of Hagedorn, momentum mode or winding
mode universes, with $\gamma=0, \pm 1/N$, respectively, in Figures
(\ref{fig4})-(\ref{fig6}), where we now also plot the flow fields
which indicate the direction of evolution for nearby trajectories.
In the plots, we start with the Eqs. (\ref{br1})-(\ref{br3}),
rescale the time variable to $d\tau=e^{(\varphi_s - \gamma N
\lambda_s)/2} dt$ and introduce the variables $l=\lp$ and $f=\vp$
to recast these equations as a first-order system,
\be f' - \half f (\gamma N l + f)=-\half E_0 \, , ~~~~~~ l' -
\half l(\gamma N l + f)= \frac{\gamma}{2}E_0 \, , ~~~~~~ f^2 -N
l^2 =E_0 \, . ~~ \label{phsp} \ee
This system has fixed points
\be \left( f, l \right)=\left(  \pm sgn(\gamma)
\sqrt{\frac{E_0}{1-\gamma^2 N}} , \pm \sqrt{\frac{\gamma^2
E_0}{(1-\gamma^2 N)}},  \right) \, , \label{fixedpts} \ee
which are precisely the relevant attractor/repeller solutions that
we discussed previously. The flow fields are sensitive to the
location of the fixed points, as controlled by the initial
comoving energy $E_0$.  However, to understand the dynamics we
must keep in mind that the flows are constrained by the last of
Eqs. (\ref{phsp}) to follow a specific trajectory in  phase space
as proscribed by the Hamiltonian constrain from the gravitational
sector, which in simpler parlance is just the Friedman equation.
An arbitrary trajectory in phase space for a fixed energy density
and pressure, that may be a solution of the first two of
(\ref{phsp}) is {\it not} a solution of the full gravity equations
unless the constraint is satisfied explicitly. So, the flow fields
in the Figures (\ref{fig4})-(\ref{fig6}) are to be understood
merely as an indicator for how the system will evolve if it is
perturbed by a source that is asymptotically subleading to the
dominant sources parameterized by $E_0$, which would deform the
system trajectory away from the depicted solid curve, with the
deformation becoming small near the fixed points. Conversely, if
we wish to follow a randomly chosen trajectory in the phase plane,
which is not one of the depicted solid lines, we {\it must}
provide the correct sources in order to ensure that the trajectory
is a solution of the full gravitational system. But in such cases,
the flow lines will be different than those depicted, particularly
if the new sources turn out to be dominant.

As a consequence of this discussion, we see immediately that if we
want to follow a trajectory that would take us to the coordinate
origin $(f,l) = (0,0)$, we must dial the value of $E_0$ to zero as
time goes on. This of course means that the fluid dominating the
universe will be neither Hagedorn, nor the momentum mode, nor the
winding mode gas. Further, as long as the energy density remains
positive, the system will have to follow a trajectory which is
composed piecewise of the infinitesimal portions of the depicted
hypebolas, because of the last of Eqs. (\ref{phsp}). This then
shows that for such evolution the coordinate origin will be an
accumulation point of the fixed points on the one-parameter family
of hyperbolas enclosed between the dashed straight lines in the
Figures (\ref{fig4})-(\ref{fig6}), and so, it itself will be a
fixed point. Therefore, it will take an infinite time to reach it.
As a result, the universe represented by such solutions will be
future geodesically complete, and its passage through the origin will be
impossible. This shows that the idea of the loitering universe
dynamics of \cite{break} cannot be realized with positive
sources. The transition from one side of the phase diagram 
to another will never complete, and the universe on one side, approaching the
fixed point, will be left hanging there forever.

One may also ask how robust are these conclusions, since quantum
corrections may violate positivity of energy, and perhaps relax some 
of the restrictions on the dynamics. 
In fact one does expect additional
contributions to the effective Lagrangian, that could arise from
other stringy matter, higher order $\alpha^\prime$ or $g_s$
corrections, spatial curvature, and various possible contributions
to the effective dilaton potential, and so this question is
relevant. If the effective energy density in
the course of evolution were negative for a while, then a change from 
$(-)$ branch solutions to $(+)$ branch
ones may occur, redirecting the flow of solutions from weak to strong
curvature regime \cite{kalmad}. However the
transitions from $(+)$ to $(-)$ branch, that could avert the
future singularity once the cosmology starts evolving towards it,
cannot happen if the sources which dominate the
universe {\it do not} violate NEC during the transition from
different string phases. To see this, we can generalize Eqs
(\ref{br1})-(\ref{br3}) to nominally include additional
contributions to the dynamics in the effective Lagrangian ${\cal
L}_{m}={\cal L}_{m} ( \phi, g_{\mu \nu}, \ldots )$ (denoted by
$\ldots$), arising from any other stringy matter, higher order
$\alpha'$ and $g_s$ corrections, spatial curvature, and various
possible corrections to the dilaton potential. Restricting the
analysis to the homogeneous and isotropic perfect fluid
cosmologies as before,
with the string ensemble sources $T_{\mu \nu}={\rm diag}\left(
\rho_s, p_s, \ldots, p_s \right)$, now including the contributions
from the corrections to ${\cal L}_{m}$ as well as
the string ensemble terms, we find \cite{brumad}
\bea
\dot{\varphi}_s&=& \pm \sqrt{N H_s^2
+ e^{\phi_s} \rho_s} \, , \label{ve1} \\
\dot{H}_s&=&\pm H_s \sqrt{N H_s^2
+e^{\phi_s} \rho_s} + \frac{1}{2} e^{\phi_s}
\left( p_s + \Delta_\phi {\cal L}_m \right) \, , \label{ve2} \\
\dot{\rho}_s&=&-N H_s \left( \rho_s + p_s \right)- \pd_s
\Delta_\phi {\cal L}_m \, , \label{ve3} \\
\vdd_s &=& N H_s^2 + \half e^{\phi_s} \left(
\rho_s - \Delta_\phi {\cal L}_m \right) \, ,
\label{helpful1} 
\eea
where $\Delta_\phi {\cal L}_m$ is the variational derivative of
the effective Lagrangian with respect to the dilaton including any couplings of the dilaton to
the general sources which are absent in the string ensemble
description (\ref{gasact}). Here clearly the upper sign refers to
the $(+)$ branch configurations, and the lower sign to the $(-)$
branch. The Eq. (\ref{helpful1}) follows by taking a derivative of (\ref{ve1}) and
straightforwardly manipulating the terms.

As noted before, a branch change may not occur unless the
discriminant, or the `egg' function \cite{bruve}
\be  \label{eggfunction} {\tt e} =N H_s^2  + e^{\phi_s}
\rho_s \, , \ee
vanishes during some stage of the evolution, which requires
$\rho_s <0$. Now, suppose that $\rho_s$ does dip below zero for 
some range of variables and parameters. From Eqs. (\ref{ve1})-(\ref{helpful1})
it is clear that the phase space trajectories can only explore the region of $\rho_s<0$
down to the contour ${\tt e}=0$. The region ${\tt e} <0$ is excluded by demanding the 
reality of the solutions. Now, the region where $\rho_s <0$ should be compact,
in order to match it asymptotically to the string gas cosmologies discussed above,
for which $\rho_s \ge 0$. Thus, the contour ${\tt e} =0$ should be bounded, with its
interior excluded, just like the `egg' region discussed in the pre-Big Bang attempts
to induce branch changing \cite{bruve,brumad}. The mechanism of branch-changing
can then be viewed as a collision between the system trajectory in the phase space and the
`egg' contour ${\tt e}=0$, that needs to stand in its way. A branch changing would then
require the following ingredients:

\bi

\item Collision with the `egg'; branch changing without collisions is impossible, since 
the sign of $\dot \varphi_s$ will not change unless $e=0$ somewhere along the trajectory.

\item At the moment of collision, $\dot \varphi_s \propto \sqrt{\tt e} = 0$. 
Since the trajectory cannot enter inside the `egg' and will not stop there, as $
\ddot \varphi_s \ne 0$, it will ricochet away, grazing the `egg'. By continuity,
the sign of $\dot \varphi_s$, and consequently the branch, will change.

\item Subsequently the trajectory should fly away from the `egg', such that further collisions
that can change the branch again will not occur. Alternatively, for the branch change to occur there 
could be only an odd number of collisions. 

\ei

So to have a branch changing, we must generate an `egg' region in the way of a trajectory
which asymptotically approaches those of Figures (\ref{fig3})-(\ref{fig6}). Then we must
tune the initial conditions such that a trajectory will glance off the `egg', and make sure that
it can flow away after an odd number of hits. 
To check when these requirements may be realized, write the
first derivative of $\vpd_s= \pm \sqrt{\tt e}$ using (\ref{helpful1}),
$\pm \frac{d}{dt} \left( \sqrt{\tt e} \right) ={\tt e} + \half
e^{\phi_s} \rho_s  - \half e^{\phi_s} \dlm$, and
using (\ref{ve2}) eliminate $\dlm$ to find
\be \pm \frac{d}{dt} \left( \sqrt{\tt e} \right) = -\dot H_s+ H_s \pd_s
+ \half e^{\phi_s} \left( \rho_s + p_s \right) \, . 
\label{diffegg}
\ee
Suppose that a branch change does occur, so that 
the trajectories arriving to the surface of the `egg' at say the instant $t_h$, when
$e=0$, and departing away from the region of the `egg' at say $t_e$ when $\rho_s = 0$ and 
so ${\tt e} = N H_s^2$, and beyond which 
again $\rho_s \ge 0$, are on different branches.
Integrating Eq. (\ref{diffegg}) between these instants yields \cite{kalmad}
\be \int_{t_h}^{t_e} dt \left[ \frac{d}{dt} \left( \pm \sqrt{\tt e} +
H_s \right) -H_s \, \pd_s \right] = \half  \int_{t_h}^{t_e} dt \;
e^{\phi_s} \left( \rho_s + p_s \right) \, .
\label{integralegg} 
\ee
The first term is a total derivative and so is given by the difference of the 
quantity $\pm \sqrt{e} + H_s$ at the limits of integration. The second quantity is
$\int^{t_e}_{t_h}  dt \, H_s \dot \phi_s = \int^{t_e}_{t_h} d\phi_s H_s$, and so it is 
given as the line integral of $H_s(\phi_s)$ over the phase space trajectory. 
Because the flow of the phase space trajectories around the `egg' is
clockwise \cite{kalmad}, which can be glimpsed from the flow lines on the 
Figures (\ref{fig4})-(\ref{fig6}), the relation $\dot \phi_s = \dot \varphi_s + N H_s$, and the 
fact that around the `egg' $H_s$ cannot change the sign since it obeys 
$NH_s^2 = {\tt e} + e^\phi_s |\rho_s| \ge e^\phi_s |\rho_s|>0$, 
this integral is {\it equal} to the area ${\cal A}$
between the projection of the phase space trajectory onto the $(\phi_s, H_s)$ hyperplane 
and hence it is positive definite. Then integrating and using that ${\tt e}(t_e) = NH_s^2(t_e)$
and $e(t_h)=0$, we get
\be 
\left(\pm \sqrt{N}-1\right) H_s(t_e) + H_s(t_h) +
{\cal A} = - \half  \int_{t_h}^{t_e} dt  \; e^{\phi_s}
\left( \rho_s + p_s \right) \, . 
\label{intcond} \ee
Note that this formula immediately applies for any odd number of hits between $t_h$ and
$t_e$, thanks to the additivity of definite integrals. 

Now, if the incoming trajectory is a $(-)$ branch expanding universe with $H_s >0$, 
keeping the lower sign in the first term of Eq. (\ref{intcond}) we can rewrite this equation 
as 
\be
\left(\sqrt{N}+1 \right) H_s(t_e) = H_s(t_h) + {\cal A} +  \half  \int_{t_h}^{t_e} dt  \; e^{\phi_s}
\left( \rho_s + p_s \right) \, ,
\label{minbrch}
\ee
which in principle allows a branch change $(-) \to (+)$ to occur even when the NEC is always
satisfied, $p_s + \rho_s \ge 0$. For such a transition, what is required is that prior to the transition,
the expansion of the universe on the $(-)$ branch, as characterized by $H_s(t_h)$, is slowed down.
This is accomplished by the effective string frame energy density $\rho_s$ dipping below zero, 
but then NEC need not be violated. Such transitions can be realized by even such mundane terms
as the positive spatial curvature of the universe, as is familiar in the examples considered in
\cite{kalmad}. However, such transitions are not desired. On the contrary, if a transition
from $(-)$ branch to a $(+)$ branch occurs, it then implies that the universe will end up in the
future singularity, unless another transition back to $(-)$ branch happens.

Those transitions, which are also required in order to realize the cosmological scenarios discussed
in \cite{tseva,break}, are much harder to accomplish. Indeed, going back to Eq. (\ref{intcond}),
to describe the $(+) \to (-)$ transition we now must keep the upper sign in the first term,
as that describes an initially $(+)$ branch solution. Then we can rewrite the resulting equation 
for an expanding universe as
\be
\left(\sqrt{N}-1 \right) H_s(t_e) + H_s(t_h) + {\cal A}  = -  \half  \int_{t_h}^{t_e} dt  \; e^{\phi_s}
\left( \rho_s + p_s \right) \, .
\label{plusbrch}
\ee
Since the universe should be expanding before and after the transition, the left hand side of this equation
is positive definite, and unless NEC is violated, we see that no solutions can satisfy it. Thus, 
if any solutions that are consistent with Eq. (\ref{plusbrch}) were to exist, NEC {\it must} be violated
in an overwhelming way - during the transition, the universe must be dominated by the sources which
do not respect NEC.

It is very difficult to imagine such sources emerging in string theory. It is true that Casimir energy violates
NEC, however in cosmological applications such contributions are by and large subleading\footnote{The 
corrections to the Hagedorn gas equation of state which have been analyzed in \cite{markh} 
indeed do not violate NEC.} 
to the dominant sources that control the evolution \cite{woodard}. One exception, where the NEC violating 
effects are studied and found to significantly affect the background is the eternal inflation \cite{eternal}. In eternal inflation, quantum fluctuations drive the scalar field up the potential, which violates NEC, and 
induces an inflationary landscape which populates the vacua of the theory. This happens
without violating the positivity of energy (while keeping the quantum matter 
of the universe in equilibrium with gravity, as opposed to just in equilibrium) 
in contrast to what branch changing processes would need to do.
On the other hand, the recent explorations of the string landscape have shown that it is possible to find inflation
in string theory \cite{kklt}, and so such dynamics may occur even if the universe starts out dominated by
some string gas. If this happens, however, inflation will generically completely wipe out the traces of the 
early initial conditions in the universe, reshaping the universe in accord with inflationary dynamics. This
would then solve cosmological problems without the need to resort to the gas phase.

A possible method to induce NEC violation is
to introduce ghosts in the theory, which of course typically lead to a host of problems with both the microscopic and the large scale behavior of physical systems, involving uncontrollable instabilities. An interesting suggestion that a ghost may condense \cite{nima}, 
and its instabilities tamed in cosmological applications
has been made in \cite{markus,paolo}, however string theory has been proven to be ghost-free
\cite{stelle,zwie}. Furthermore, the ghost condensate model \cite{nima} is known to have a very low cutoff, which is necessary to keep the bad behavior under control. To go beyond the cutoff and explore
energy scales such as those encountered in string cosmology, one should need to find a UV completion of the theory and use it at the scales above the cutoff. The investigation of the possible completions however
has turned up arguments against such completions based on locality and causality \cite{nimauv}. 
Thus unless it is demonstrated that the ghost condensate can be consistently embedded in string theory,
the interesting condensed ghost bounces of \cite{markus} do not really provide the
mechanism to generate branch changes in string gas cosmologies. Thus for now, it appears that 
the phenomenological applications of  $(+)$ branch solutions, that evolve towards future singularities, invariably leads one into the string swampland \cite{swamp}.

\section{Summary}

To summarize, in this paper we have considered the 
obstructions to phase transitions between different thermodynamic phases of the string gas
in the early universe. These arise because the string dilaton and gravity react to the background 
sources by generating the flow of the dilaton, or the string coupling. If the 
energy density of the sources is non-negative, the sign of the time derivative of the
dimensionally reduced dilaton $\varphi_s = \phi_s - N \lambda_s$ is conserved, behaving like 
a discrete charge, and separating the solutions into two superselection sectors,
or $(+)$ and $(-)$ branches. The $(+)$ branch solutions are problematic since they
evolve to a future spacelike singularity, which looks either like a Big Crunch or a pole inflation
in the string frame. The singularity separates different branches of string gas solutions like
a geometric precipice in the moduli space, and prevents them from joining together.
The flow cannot be altered without negative energy and NEC violations. 
This excludes the early universe cosmologies based on the transitions of the string Hagedorn
gas evolving away from the fixed point to the FRW radiation cosmology \cite{tseva}, 
and the loitering gas models of \cite{break}, unless new sources are introduced which
violate both positivity of energy and NEC. We provide quantitative criteria which such sources must 
satisfy to yield a branch change, but recall that the known mechanisms which could in principle satisfy these
criteria \cite{markus,paolo} do not at present come out of string theory in a natural way.
It does remain possible to build composite cosmologies using the solutions that reside on the same branch, 
and we have seen an example of a model patched together from winding mode, Hagedorn and momentum mode cosmologies on the $(-)$ branch. Those solutions behave differently from the ones discussed in
\cite{tseva,ali} because they do not have a future singularity. The past singularity could be `resolved' by using $T$-duality, instead of bounces induced by negative energy and NEC violations, and is closer in spirit to the original ideas of how $T$-duality might help in cosmological applications. These solutions do not yet
have interesting cosmological applications, but it would be interesting to consider what happens with cosmological perturbations in universes that are $T$-dual to each other.

\vskip1.5cm

{\bf \noindent Acknowledgements}

\smallskip

We thank A. Linde and M. Johnson for useful discussions, and 
the Galileo Galilei Institute for Theoretical
Physics in Firenze, Italy, for kind hospitality during the
inception of this work. SW thanks the UC Davis HEFTI program for
support and hospitality during the course of this work. 
SW would also like to acknowledge Ruby Matthews 
for a lifetime of inspiration.
The research of NK is supported in part by the DOE Grant
DE-FG03-91ER40674 and in part by a Research Innovation Award from
the Research Corporation.


\end{document}